\documentclass[runningheads]{llncs}
\usepackage{graphicx}
\usepackage{multirow}
\usepackage{booktabs}
\usepackage{siunitx}

\begin{document}
\title{Evaluating X-vector-based Speaker Anonymization under White-box Assessment}
\titlerunning{White-box Assessment of Speaker Anonymization}
%
\author{Pierre Champion\inst{1,2} \and
Denis Jouvet\inst{1} \and
Anthony Larcher\inst{2}}
\authorrunning{P. Champion et al.}
%
\institute{Université de Lorraine, CNRS, Inria, LORIA, F-54000 Nancy, France
\email{\{pierre.champion, denis.jouvet\}@inria.fr}\\
 \and
Le Mans Université, LIUM, France\\
\email{anthony.larcher@univ-lemans.fr}}
\maketitle              
\begin{abstract}


In the scenario of the Voice Privacy challenge, anonymization is achieved by converting all utterances from a source speaker to match the same target identity; this identity being randomly selected.
In this context, an attacker with maximum knowledge about the anonymization system can not infer the target identity. 
This article proposed to constrain the target selection to a specific identity, i.e., removing the random selection of identity, to evaluate the extreme threat under a white-box assessment (the attacker has complete knowledge about the system).
Targeting a unique identity also allows us to investigate whether some target's identities are better than others to anonymize a given speaker.

\end{abstract}
\setcounter{footnote}{0}

\section{Introduction}

In many applications, such as virtual assistants, speech signal is sent from the 
user device to the service provider's servers in which data is collected, processed, and stored. 
Recent regulations, e.g., the General Data Protection Regulation (GDPR) \cite{gdpr} in the EU, 
emphasize on privacy preservation and protection of personal data.
As speech data can reflect both biological and behavioral characteristics 
of the speaker, it is qualified as personal data \cite{nautschGDPRSpeechData2019}.
The research reported in this article has been done using the Voice privacy Challenge framework \cite{tomashenkoVoicePrivacy2020Challenge}, which is one of the first attempts of the speech community to evaluate research on this topic,
by producing dedicated protocols, metrics, datasets and baselines.

Speaker anonymization is performed to suppress the personally identifiable paralinguistic information from a speech utterance while maintaining the linguistic content.
This is also referred to as \textit{speaker 
anonymization}\cite{fangSpeakerAnonymizationUsing2019} 
or \textit{de-identification} \cite{magarinos2017reversible}.
Recently, Fang et al. \cite{fangSpeakerAnonymizationUsing2019} proposed an x-vector-based speaker anonymization system based on voice conversion where one of the hyper-parameters used to transform the voice is a target pseudo-speaker identity. In order to choose the anonymization system and hyper-parameters, for a given use case, we must evaluate and rank the performances of each anonymization method. 
In this first edition of the Voice privacy Challenge, the quality of anonymization is assessed by using state-of-the-art speaker verification system together with an automatic speech recognition system that is used to evaluate the preservation of the linguistic content.
In the context of privacy, a game-theoretic reasoning with two agents is defined as follows: one who wishes to anonymize a user's speech (the service provider) and the other (the attacker) who will do everything possible to "break" this anonymity.
Introduced in \cite{EvaluatingVoiceConversionbased2019}, the notion of attacker's knowledge of the anonymization method was introduced, it defines three possible attack scenarios: black-box, white-box, and grey-box.
In all approaches, privacy is evaluated by comparing original speech (accessible to the attacker) and anonymous speech published by a service provider.
The goal of an effective anonymization system is to allow the publication of speech whose identity is difficult to link to a user's identity, even if the attacker has some knowledge of the anonymization mechanism.
In the black-box attacker scenario, privacy is measured by comparing clean, non-anonymized enrollment speech, and anonymized trial speech. In this scenario, the attacker is not aware that speech has been anonymized.
In contrast, a white-box attacker is fully aware of the anonymization system and hyper-parameters. 
Grey-box attackers cover the whole range of possible scenario in-between black and white boxes, when the attacker has a partial knowledge of the anonymization system.

The target pseudo-speaker identity used to anonymize the voices can be generated following multiple strategies \cite{EvaluatingVoiceConversionbased2019}. 
In the \textit{permanent} strategy, used in the VoicePrivacy challenge, all utterances from a speaker are converted using the same target pseudo-speaker. 
The challenge defines a grey-box attacker that has access to the anonymization toolkit, and thus is able to transform any utterance from her enrollment dataset using the same system and target strategy. 
However, for a given source speaker, the target pseudo-speaker used by the attacker differs from the one used by the service provider due to the random pseudo-speaker selection process included in the \textit{permanent} strategy.
In case the attacker knows the pseudo-speaker, this grey-box scenario turns into a white-box scenario.

We modify the Voice Privacy Challenge scenario into a white-box assessment by targeting a same speaker identity for all utterances from all speakers.
This target selection strategy allows us to evaluate the performance of the x-vector-based anonymization system on its own, and generate a report that does not assume that the attacker has knowledge deficiencies.
%
Experiments are performed with a large group of target speaker identity in order to investigate the effect of this identity on the quality of the anonymization.

In the remaining of this article, we first describe the baseline system and voice conversion method in Section \ref{sec:baseline}.
We then introduce our experimental protocol in Section \ref{sec:experiments} and present our experimental
results in Section \ref{sec:expe}. Eventually, we draw our conclusions and propose future avenues in Section \ref{sec:conclusion}.

\section{Anonymization Technique}

\label{sec:baseline}

The anonymization of speaker's identity can be performed with various methods \cite{F0_Bahmaninezhad2018ConvolutionalNN,fangSpeakerAnonymizationUsing2019,mcadams,mohanPrivacyPreservingAdversarialRepresentation2019,EvaluatingVoiceConversionbased2019}. In this article, our contributions are based on the \textit{Baseline-1} (referred to as the \textit{baseline} in this article) of the VoicePrivacy challenge that anonymizes speech using x-vectors and neural waveform models \cite{fangSpeakerAnonymizationUsing2019}.  

\subsection{The Voice Conversion System}

\vspace{-1em}

\begin{figure}[ht]
  \centering
  \includegraphics[width=0.9\linewidth]{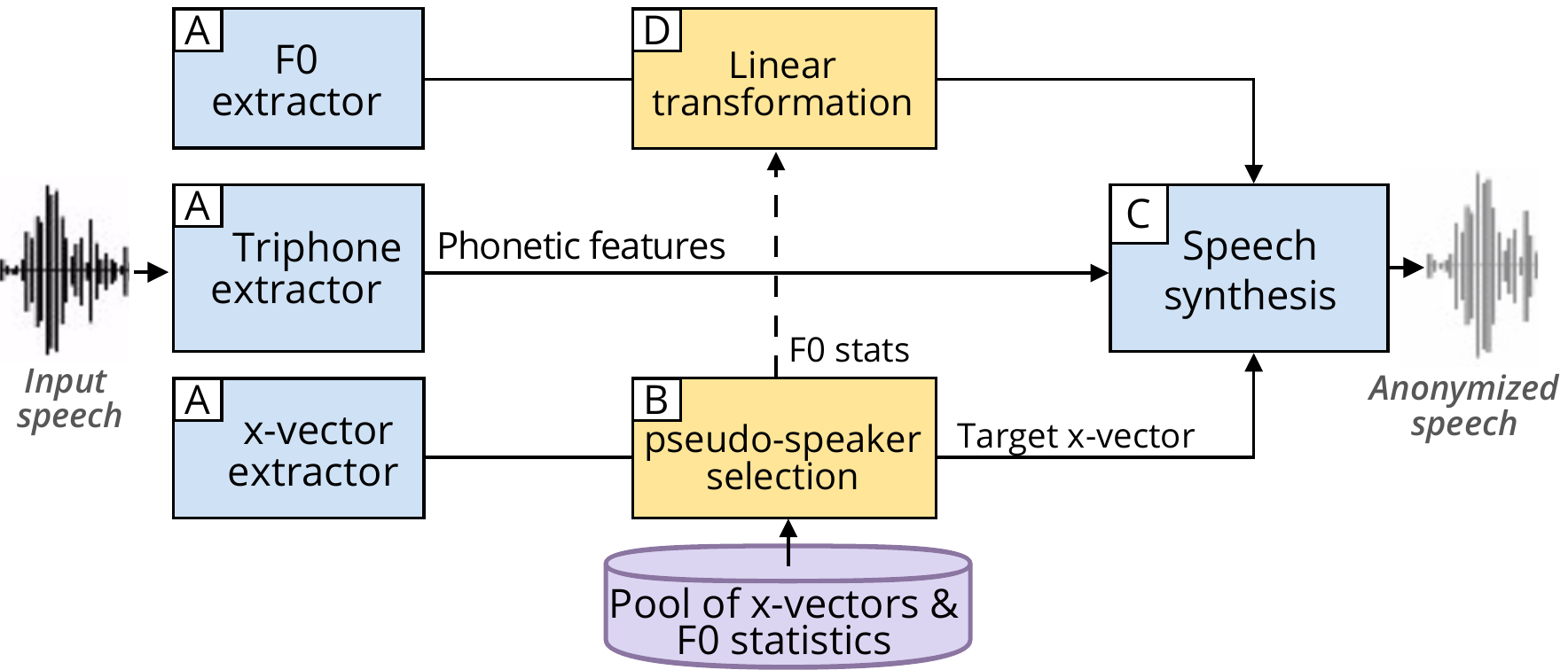}
  \caption{The speaker anonymization pipeline. Modules A, B, and C are parts of the baseline model. Module D is an enhancement used to coherently alter the F0 values with respect to the selected pseudo-speaker identity.}
  \label{fig:baseline.png}
\end{figure}

\vspace{-1em}

The baseline system introduced in \cite{fangSpeakerAnonymizationUsing2019} aims at separating speaker identity and linguistic content from an input speech utterance. 
Assuming that those features are disentangled, an anonymized speech waveform can be generated after altering only the features that encode the speaker's identity.
The anonymization system illustrated in {Figure \ref{fig:baseline.png}} can be decomposed into three groups of modules.
Modules from the \textit{group A} extract different features from the source signal: the fundamental frequency, the phonetic features encoding articulation of speech sounds (Phoneme Posterior-Grams (PPGs) \cite{ppgs}) and the speaker's x-vector.
\textit{The module B} derives a new pseudo-speaker identity. The x-vector from each source input speaker is compared to a pool of external x-vectors to select the 200 furthest vectors; 100 of them are randomly selected and averaged to create an anonymized pseudo-speaker x-vector identity. 
Finally, \textit{the module C} synthesizes a speech waveform from the pseudo-speaker x-vector together with the original PPGs features and F0.

As an enhancement to this baseline, \cite{f0_mod_moi} proposed to modify the F0 values (cf. module D in Figure \ref{fig:baseline.png}) using the F0 mean and variance statistics associated with each of the x-vector of the pool.

\subsection{Design Choices for Anonymization}

\begin{figure*}[!hb]
\hspace{-0.16cm}
\begin{minipage}[b]{0.47\linewidth}
  \centering
  \centerline{\includegraphics[width=1.0\linewidth]{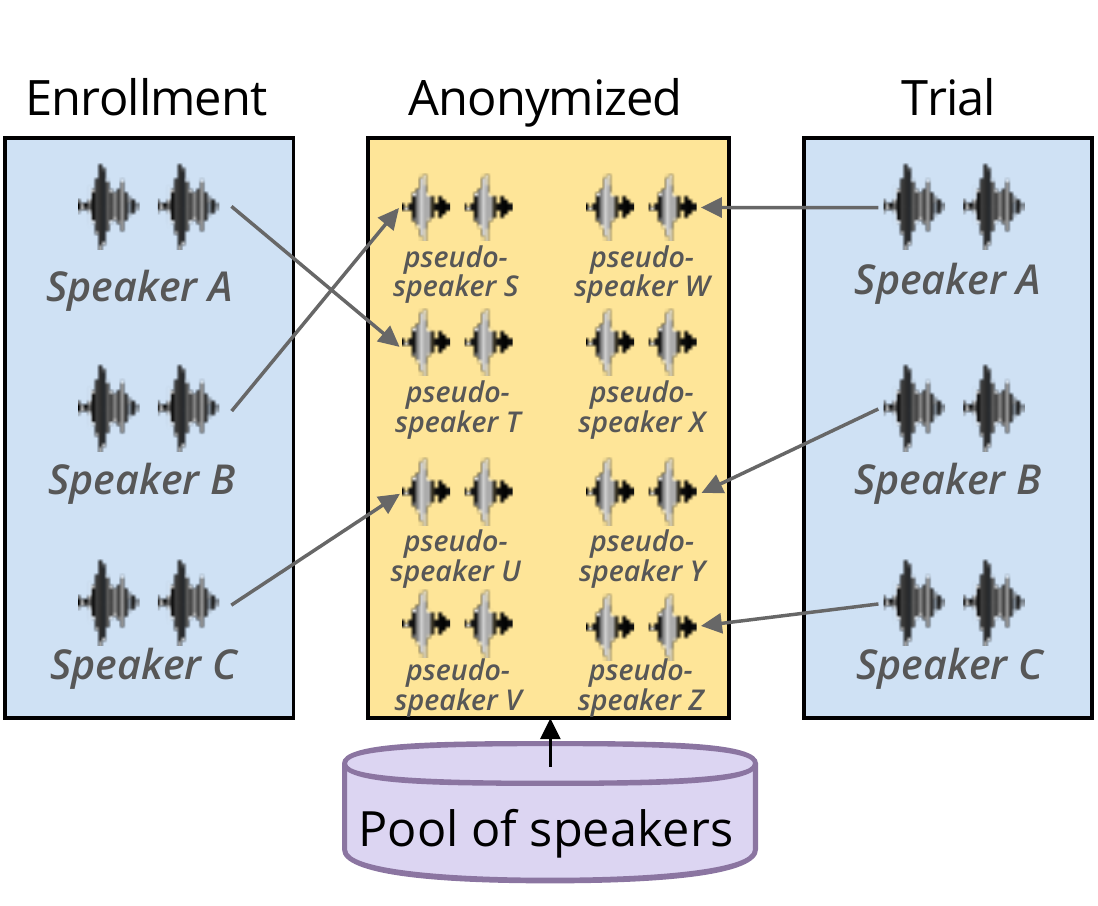}}
  \vspace{0.2cm}
  \centerline{\begin{minipage}{1.5in}\centering a) VoicePrivacy using the \textit{permanent} strategy\end{minipage}}
\end{minipage}
\hfill
\hspace{0.2cm}
\begin{minipage}[b]{0.47\linewidth}
  \centering
  \centerline{\includegraphics[width=1.0\linewidth]{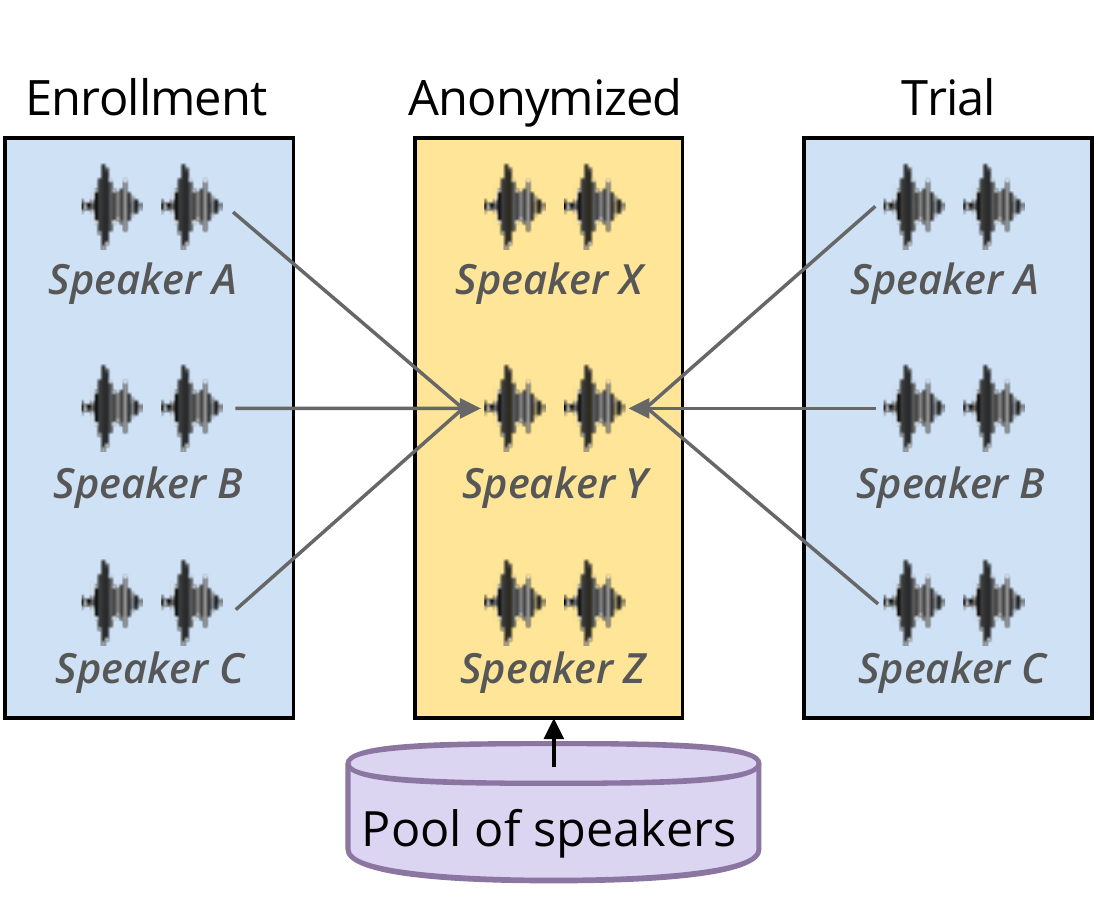}}
  \vspace{0.2cm}
    \centerline{\begin{minipage}{1.7in}\centering b) Our evaluation using the \textit{constant}
  strategy\end{minipage}}
\end{minipage}
\caption{Overview of target selection strategies. In the \textit{permanent} strategy pseudo-speakers are randomly selected, first for the trial dataset and then for the enrollment dataset. While our \textit{constant} strategy, always targets the same pseudo-speaker.}
\label{fig:target_selection}
\end{figure*}

In the evaluation plan of the VoicePrivacy Challenge, all utterances of a given speaker should be converted to match a single target pseudo-speaker. This strategy, described as \textit{permanent} in \cite{EvaluatingVoiceConversionbased2019}, ensures that a one-to-one mapping exists between the source speaker identity and the anonymized speaker identity. 
This requirement allows anonymized voices to be distinguishable from each other.
%
This one-to-one mapping does not apply in-between speech anonymized by the service provider and speech anonymized by the attacker.
Speech data anonymized by the service provider corresponds to the trial dataset, and speech available to the attacker corresponds to the enrollment dataset, see Section \ref{data} for more detail.
Figure \ref{fig:target_selection}a shows how the \textit{permanent} strategy converts enrollment and trial speech to different anonymized pseudo-speakers.

To select the target pseudo-speaker identity, module B (from Figure \ref{fig:baseline.png}) has many hyper-parameters that affect the selection mechanism.  
According to \cite{f0_mod_moi,brij_journal,brij_vpc_design}, the best anonymization results are achieved by picking the pseudo-speaker in a dense region of the x-vector space, randomly targeting male or female gender pseudo-speaker, and modifying the F0 values of the input speech so that it matches the F0 statistics of the real speakers used to generate the pseudo-identity. 


The VoicePrivacy challenge's protocol assumes that attackers have access to anonymized trial utterances and original enrollment utterances. 
During the challenge, two sets of tests are performed following black-box and grey-box attacker scenarios 
corresponding to situations where the enrollment utterances are original or have been anonymized. 
In the latter scenario, the attacker has partial knowledge of the system and is able to anonymize the enrollment utterances using the same anonymization system and \textit{permanent} target selection strategy. 
The pseudo-speaker chosen for each of the enrollment speakers differs from the pseudo-speaker chosen for the trial speakers as the attacker does not have knowledge of the randomly selected speakers used to generate the pseudo-speaker identity.
As the voices of the same speakers once anonymized by the service provider (trial) and the attacker (enrollment) are different from each other, this leads to a rather good anonymization performance \cite{tomashenkoVoicePrivacy2020Challenge}.

With the \textit{permanent} target selection strategy used in the challenge, anonymized voices remain distinguishable and all utterances from the same original speaker are anonymized with the same pseudo-speaker. This process is referred to as pseudonymization.
Providing attackers the target's identity of each speaker in the \textit{permanent} strategy evaluates the quality of the voice conversion system and the preservation of the voice distinctiveness \cite{similarity_matrices}.
In this paper, we evaluate the privacy of the anonymization technique with the best attacker, that is using the \textit{constant} selection strategy.


\section{Experimental Setup}

\label{sec:experiments}

In this evaluation, we want to provide a comprehensive assessment of the voice conversion toolkit under a white-box scenario where the attacker has full knowledge about the system. 
We change the game between the attacker and the service provider to use the \textit{constant} identity selection strategy defined in \cite{EvaluatingVoiceConversionbased2019}. 
In contrast to the anonymization performed in the VoicePrivacy challenge, the \textit{constant} strategy defines a single pseudo-speaker identity for all speakers in a given dataset.
We extend this property to all speakers of the trial and the enrollment datasets so that all of them should have the same anonymized voice identity.
We expect good anonymization performance as the voices of all speakers should appear to be spoken by a single identity. This breaks the 
one-to-one requirement of the challenge, speakers will not be distinguishable from each other.
Still, we believe this assessment is complementary to the realistic attacker-based evaluation of the challenge.
Figure \ref{fig:target_selection} shows the differences in the target selection strategies between the VoicePrivacy strategy (\textit{permanent} strategy), and the strategy chosen for this study (our \textit{constant} strategy).
Experiments are performed with different target speaker identities to provide averaged global results about the voice conversion toolkit, and detailed analysis of the target identity effect on multiple source speakers.

\subsection{Dataset}

\label{data}

All evaluation tasks for the experiments follow the conditions presented on the publicly available VoicePrivacy challenge\footnote{\url{https://github.com/Voice-Privacy-Challenge}}.
The triphone extractor has been trained on the \textit{train-clean-100} and \textit{train-other-500} subsets of LibriSpeech. The x-vector extractor has been trained on VoxCeleb-1,2. The speech synthesis system has been trained on the \textit{train-clean-100} subset of LibriTTS. Finally, the \textit{train-other-500} subset of LibriTTS has been used to create a pool of x-vector and F0 statistics.
The evaluation dataset is built from LibriSpeech \textit{test-clean}.
Details about the number of speakers and utterances in the enrollment and trial dataset are reported in Table \ref{tab_data_asv}.

\begin{table}
\caption{Statistics of the evaluation dataset~\cite{tomashenkoVoicePrivacy2020Challenge}.}\label{tab_data_asv}
\centering
\begin{tabular}{c@{\hskip 0.15in}l@{\hskip 0.15in}r@{\hskip 0.1in}r@{\hskip 0.1in}r@{\hskip 0.1in}}
\toprule
  \textbf{Subset} & & \textbf{Female} & \textbf{Male} & \textbf{Total} \\\midrule
  \multirow{3}{*}[-2mm] { \parbox{1.6cm}{\centering Librispeech test-clean}} & Speakers in enrollment & 16 & 13 & 29

\\ \cline { 2 - 5 } & Speakers in trials & 20 & 20 & 40
\\ \cline { 2 - 5 } & Enrollment utterances & 254 & 184 & 438
\\ \cline { 2 - 5 } & Trial utterances & 734 & 762 & 1496
\\
\bottomrule
\end{tabular}
\end{table}


\subsection{Utility and Privacy Metrics}

To evaluate the performance of the system in both privacy (\textit{speaker's concealing} capability) and utility (\textit{content intelligibility}) two systems and metrics are used.
To quantitatively evaluate the privacy, an x-vector-PLDA based Automatic Speaker Verification (ASV) architecture provided by the challenge organizers is used. The privacy protection is measured with the linkability metric: \cite{linkability} introduced two different measures that are calculated based on mated and non-mated likelihood score distributions. 
In this case, mated scores are computed comparing anonymized speech of the same user. Whereas non-mated scores are computed comparing anonymized speech of different users.
The local measure $D_{\leftrightarrow}$ is the local score-wise measure depending upon the likelihood ratio between the mated and non-mated sample's score.
The global linkability measure $D_{\leftrightarrow}^{\mathrm{sys}}$ is the average value of $D_{\leftrightarrow}$ over all mated scores. 
To obtain the linkability score for a given speaker, the average is taken from all mated scores of this specific speaker.
Work in \cite{comparative_metric} advocate the use of $D_{\leftrightarrow}^{\mathrm{sys}}$ as a robust privacy metric.
The lower the $D_{\leftrightarrow}^{\mathrm{sys}}$, the better the speakers are anonymized.
As the Equal Error Rate (EER$_\%$) measure is more often used in speaker verification, we present our result in terms of both EER$_\%$ and $D_{\leftrightarrow}^{\mathrm{sys}}$.
The higher the EER$_\%$, the better the speakers are anonymized.

For the utility, the pre-trained Automatic Speech Recognition (ASR) system provided by the challenge organizers is used to decode the anonymized speech and compute the Word Error Rate (WER$_\%$). In this evaluation, the WER$_\%$ measure is used to evaluate how the content is kept intelligible.
The lower the WER$_\%$ is, the more intelligible the anonymized speech is.
Both ASR and ASV systems are trained on LibriSpeech \textit{train-clean-360} using Kaldi \cite{KaldiPovey}. 

\subsection{Evaluation Methodology}

\label{protocol}

In this experiment, we run the anonymization and evaluation on 40 real target speaker's identities that cover as best as possible the speaker space. Thus, to select the target speakers, we identify 20 female and 20 male clusters of x-vectors in the anonymization pool (\textit{LibriTTS train-other-500}) using K-Means. We then pick the speaker x-vector that is the closest to the centroid of each cluster. The assessment of the anonymization performances will be done for each of those 40 target speaker identities. Real speaker identities are used instead of pseudo-speaker identities derived from multiples speakers. 

To evaluate the performances of the anonymization system, we perform the following procedure for the 40 selected target speaker identities.
First, we convert all utterances of \textit{Librispeech test-clean} (data corresponding to enrollment and trial) and \textit{Librispeech train-360} to the selected target speaker.
Then, we train the ASV model on the anonymized \textit{Librispeech train-360} dataset. 
Lastly, we evaluate the privacy performances for each of the speakers of \textit{Librispeech test-clean} using the specially trained ASV model.
As a result, we obtain one score for each trial speaker of \textit{Librispeech test-clean} and target identities (that is a total of $29\times40$ scores). 
To evaluate the quality of the conversion process in terms of utility (linked to speech recognition performance), we use a pre-trained ASR model released for the VoicePrivacy Challenge.
To study the effectiveness of the F0 anonymization (module D of Figure \ref{fig:baseline.png}), we perform the anonymization process, and the associated evaluation, with and without the F0 transformation enabled.



\section{Experimental Results}

\label{sec:expe}

\subsection{Global Results}

Table \ref{table:all_eer_dsys} compares the anonymization performance on a global scale.
The first line presents the linkability when no anonymization is performed (i.e., on original speech data), clean speech encapsulates the speaker's information to a high degree ($D_{\leftrightarrow}^{\mathrm{sys}}$ scores $>$ 0.90). 
The second and third lines display the linkability when using the voice anonymization system without and with the F0 transformation. 
\begin{table}
  \centering
  \caption{Linkability ($D_{\leftrightarrow}^{\mathrm{sys}}$) and EER$\%$ scores 
  for original and anonymized speech. For the anonymized data, the mean and standard deviation values are calculated over the 40 experiments (i.e., one for each target speaker). Results on anonymized data are given without and with F0 transformation.
  }\label{table:all_eer_dsys}
  \begin{tabular}{
    l@{\hskip 0.2in}@{\extracolsep{0.1in}}
    S[table-format=1.2]@{\,\( \pm \)\,}@{\extracolsep{0.03in}}
    S[table-format=1.2]@{\extracolsep{0.2in}}
    S[table-format=2.1]@{\,\( \pm \)\,}@{\extracolsep{0.03in}}
    S[table-format=1.1]@{\extracolsep{0.4in}}
    S[table-format=1.2]@{\,\( \pm \)\,}@{\extracolsep{0.03in}}
    S[table-format=1.2]@{\extracolsep{0.2in}}
    S[table-format=2.1]@{\,\( \pm \)\,}@{\extracolsep{0.03in}}
    S[table-format=1.1]@{\extracolsep{0.4in}}
    }
    \toprule
   & \multicolumn{4}{c}{Female speakers} & \multicolumn{4}{c}{Male speakers}  \\[0.05in]
    &  \multicolumn{2}{c}{$D_{\leftrightarrow}^{\mathrm{sys}}$} & \multicolumn{2}{c}{EER$\%$}   &  \multicolumn{2}{c}{$D_{\leftrightarrow}^{\mathrm{sys}}$} & \multicolumn{2}{c}{EER$\%$} \\
    \midrule
    Original     & \multicolumn{1}{c}{\hspace{-4mm}0.90} &  & \multicolumn{1}{c}{\hspace{-0.5mm}7.66} &  & \multicolumn{1}{c}{\hspace{-4mm}0.96} & & \multicolumn{1}{c}{\hspace{-0.5mm}1.11}  &  \\
    Anon. without F0  & 0.72 & 0.01 & 12.1 & 0.6 & 0.77 & 0.01 & 9.5 & 0.6 \\
    Anon. with F0   & 0.74 & 0.01 & 11.6 & 0.6 & 0.75 & 0.01 & 10.6 & 0.8 \\
    \bottomrule
  \end{tabular}
\end{table}

In contrast with the original speech results, the anonymized results come from 40 ASV tests, each using a different speaker identity. 
The mean and standard deviation values are calculated from the 40 evaluations.
From the linkability score difference between original and Anon. lines of table \ref{table:all_eer_dsys} we can conclude that speakers are less linkable to their true identity after applying the x-vector base anonymization system. From the original data to the anonymized speech, linkability scores drop by at least 17\%, meaning the anonymization system has some effectiveness.
The comparison between the last two lines shows that modifying the F0 values does not help to remove speaker information from the speech signal, as shown by the standard deviation values. 
This means that the transformation applied was not strong enough to actually remove extra speaker information or that the ASV system is quite robust against such modification.
The rather low standard deviation values across all scores show that there is no large variation when changing target speaker identity, this indicates that a given target speaker isn't more suited than another to anonymize the whole dataset. 
As there are not large scores differences between voice conversion without or with F0 values transformation, the following sections will only analyze the results of the anonymization system that includes the F0 linear transformation.

\subsection{Detailed Analysis}

\vspace{-1em}

\begin{figure*}[!h]
\begin{minipage}[b]{1\linewidth}
\vspace{-0.2cm}
  \centering
  \centerline{\includegraphics[width=0.6\linewidth]{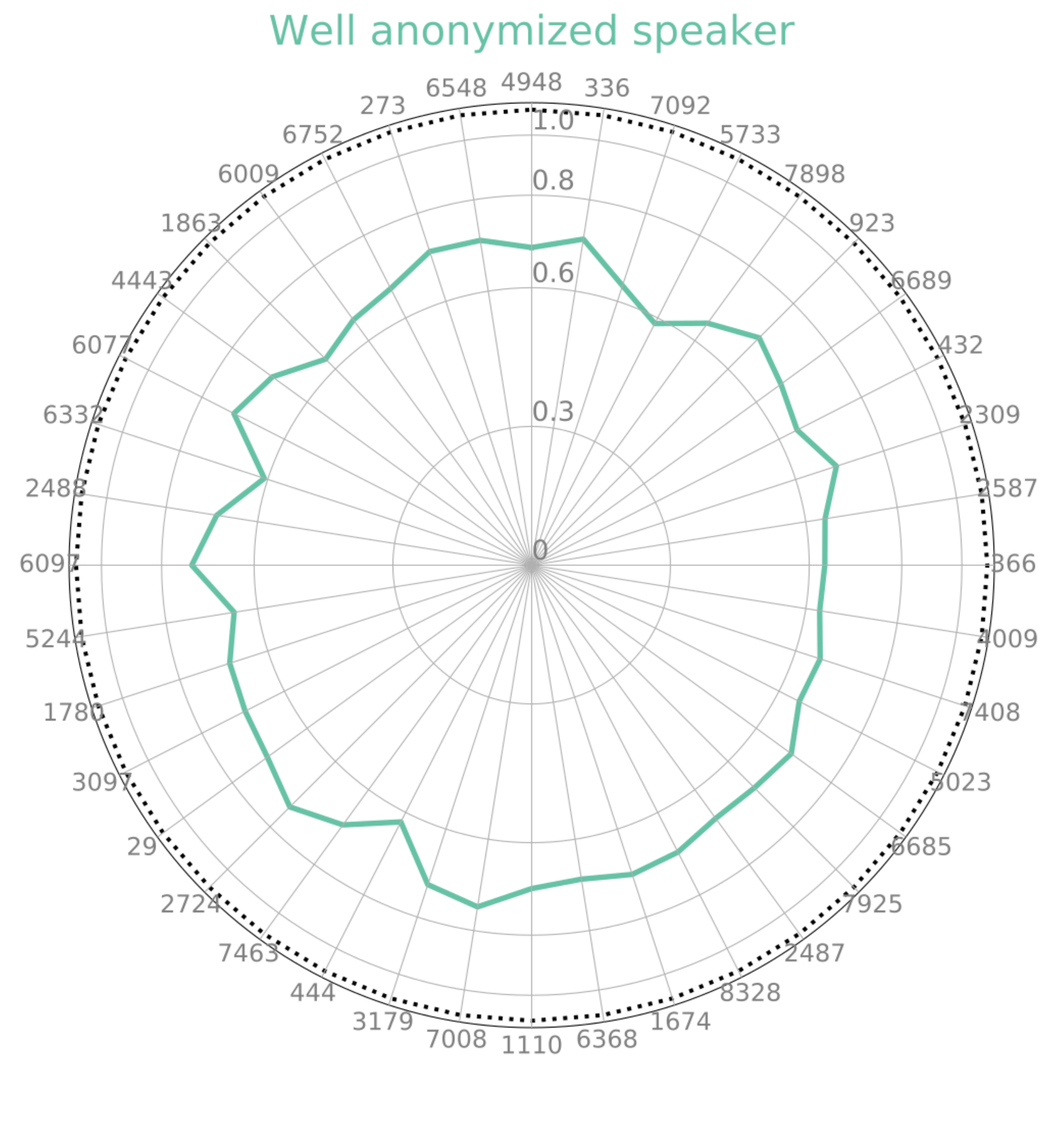}}
  \vspace{0.2cm}
  \centerline{Results on a single test speaker (speaker 5105)}
\end{minipage}
\caption{Linkability scores ($D_{\leftrightarrow}^{\mathrm{sys}}$) obtained using the ASV retrained on anonymized data for each of the 40 target speakers (colored solid line) and on original speech (dotted black circle). Each axis corresponds to a target speaker. The anonymization system lowered the linkability scores, meaning better privacy is achieved.}
\label{fig:ASV40-res}
\vspace{-0.2cm}
\end{figure*}

We conducted a detailed analysis to check whether a specific target identity is more suited to anonymize one or more speakers of our test dataset. 
Figure \ref{fig:ASV40-res}~ illustrate the visualization used for this study in the case of a single source speaker. The linkability $D_{\leftrightarrow}^{\mathrm{sys}}$ scores are computed for speech anonymized with 40 different x-vectors and F0 statistics. 

The black-dot circle indicates the linkability of the speaker on clean original speech signal (note that the original speech does not depend on the target speakers, hence the circle). And, for each of the 40 target speakers, the linkability is presented by the colored solid line.
After transforming the speech with 40 target speakers, we can observe that none of the 40 targets are significantly better to anonymize this speaker's voice. 
The variation between the anonymized linkability scores is more likely due to the difference between the ASV model training rather than a better target choice.
This observation also applies to the 29 other speakers of our test dataset.
It is also noteworthy that, out of the 40 target identities, 20 of them induce cross-gender voice conversion. 
Results for same-gender and cross-gender voice anonymization were found similar.

\vspace{-1em}

\begin{figure*}[!h]
\begin{minipage}[b]{0.5\linewidth}
\vspace{-0.2cm}
  \centering
  \centerline{\includegraphics[width=1.0\linewidth]{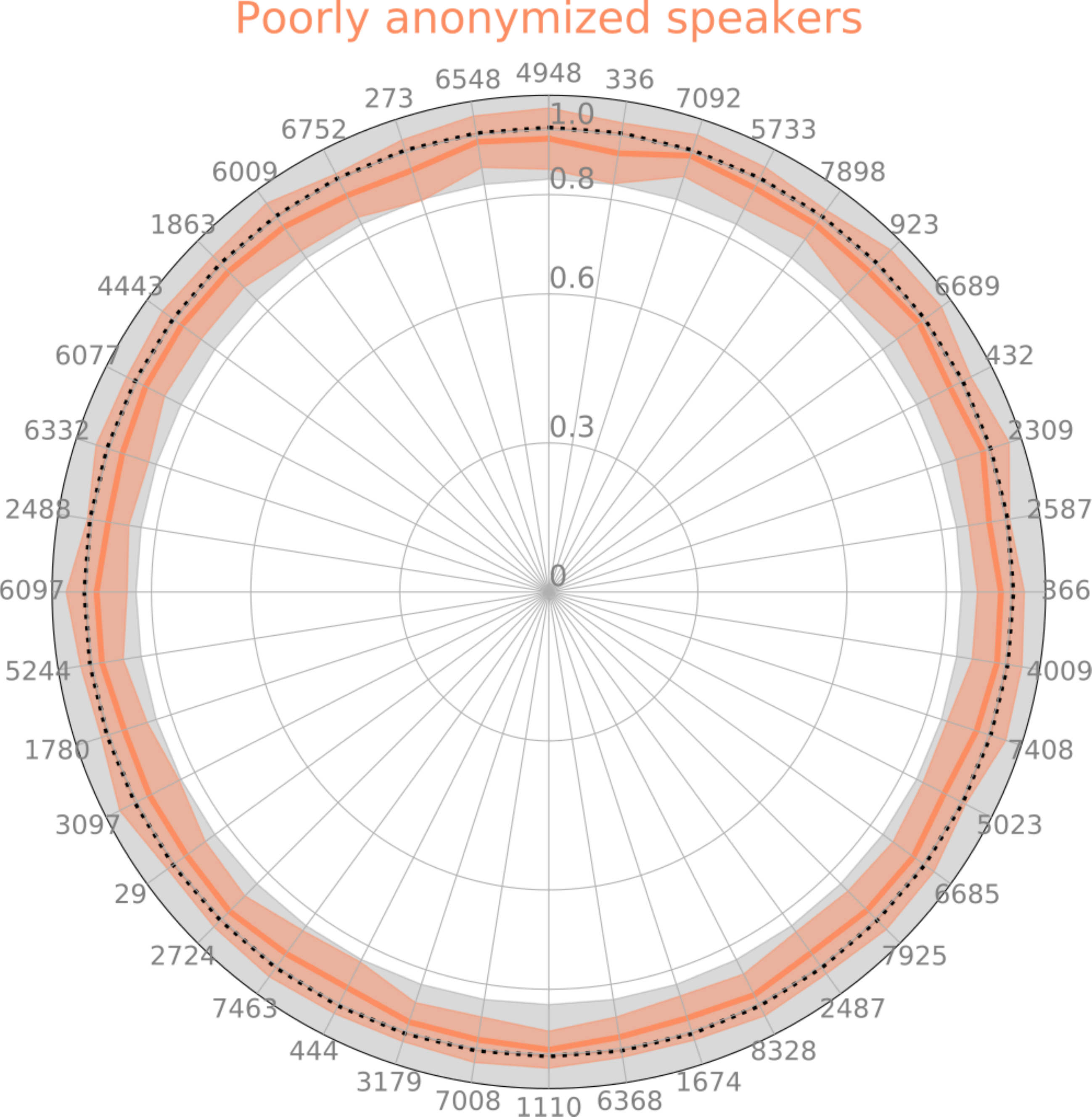}}
  \vspace{0.2cm}
  \centerline{Statistics from $N=15$ speakers}
\end{minipage}
\hfill
\begin{minipage}[b]{0.5\linewidth}
  \centering
  \centerline{\includegraphics[width=1.0\linewidth]{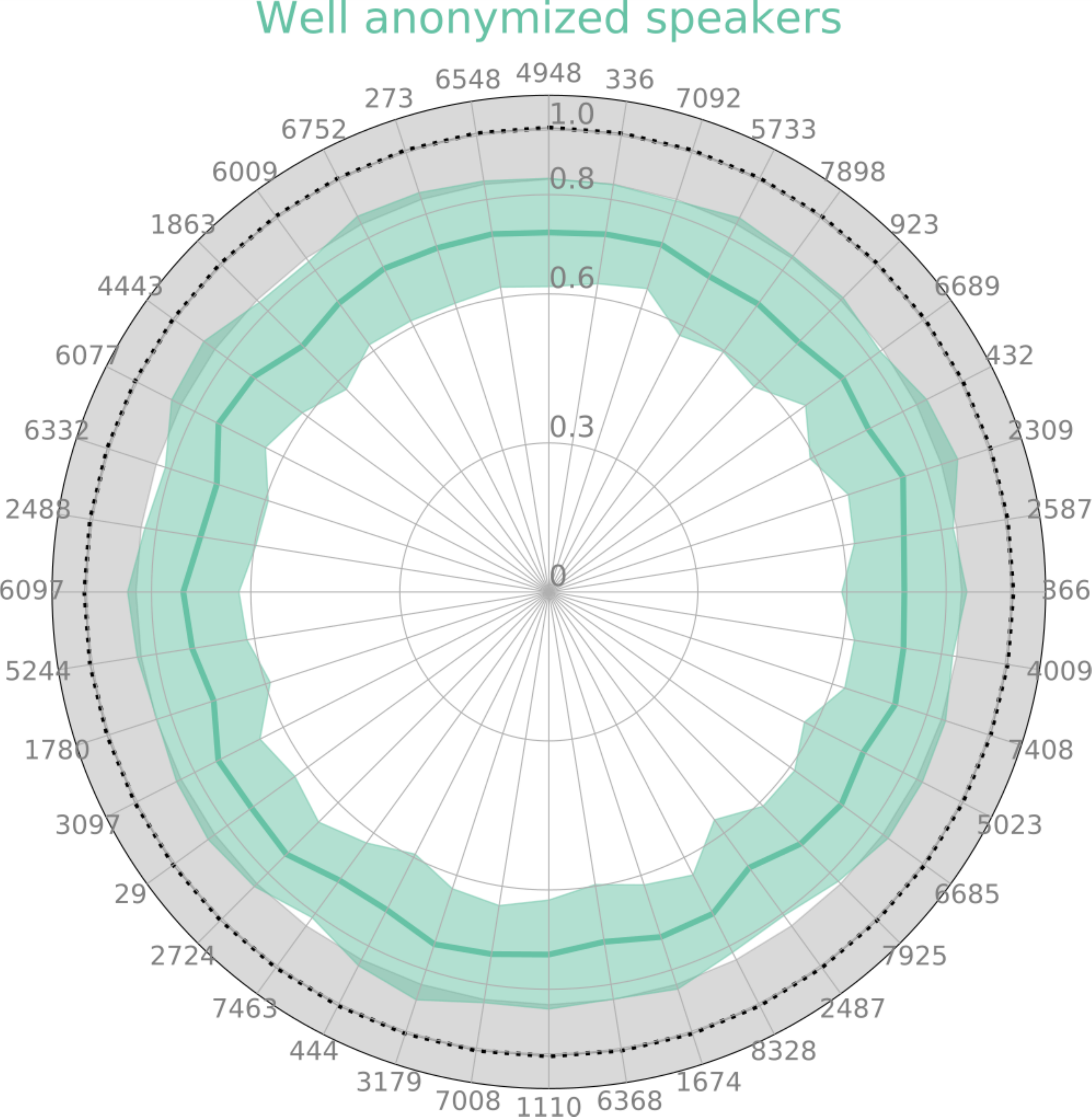}}
  \vspace{0.2cm}
  \centerline{Statistics from $N=14$ speakers}
\end{minipage}

\caption{Mean and standard deviation $D_{\leftrightarrow}^{\mathrm{sys}}$ scores obtained for $N$ speakers using retrained ASVs on anonymized data for each of the 40 target speakers (mean and standard deviation corresponding to the colored solid line and light area) and on original speech (dotted black circle and grey area). Each axis corresponds to a target speaker.}
\label{fig:ASV40-res-more}
\end{figure*}

\vspace{-1em}

Figure \ref{fig:ASV40-res-more} shows the linkability $D_{\leftrightarrow}^{\mathrm{sys}}$ scores of two groups to illustrate the common anonymization behavior: one for which the anonymization system did not remove speaker information, and the other for which the anonymization did remove some speaker information. 
For the poorly anonymized speakers, we observe that the distributions of linkability scores on anonymized speech (colored area) completely overlaps the distribution of linkability scores on original speech (grey area). 
The anonymization system did not remove any speaker information for half of our test speakers. 
On the other hand, for the well anonymized speakers, the anonymized speech and original speech scores distributions diverge. 
The difference is distinct, speaker information was removed by the anonymization system for the other half of our test speakers.

\subsection{Utility Results}
\begin{figure*}[!ht]
\begin{minipage}[b]{1\linewidth}

  \centering
  \centerline{\includegraphics[width=1.2\linewidth]{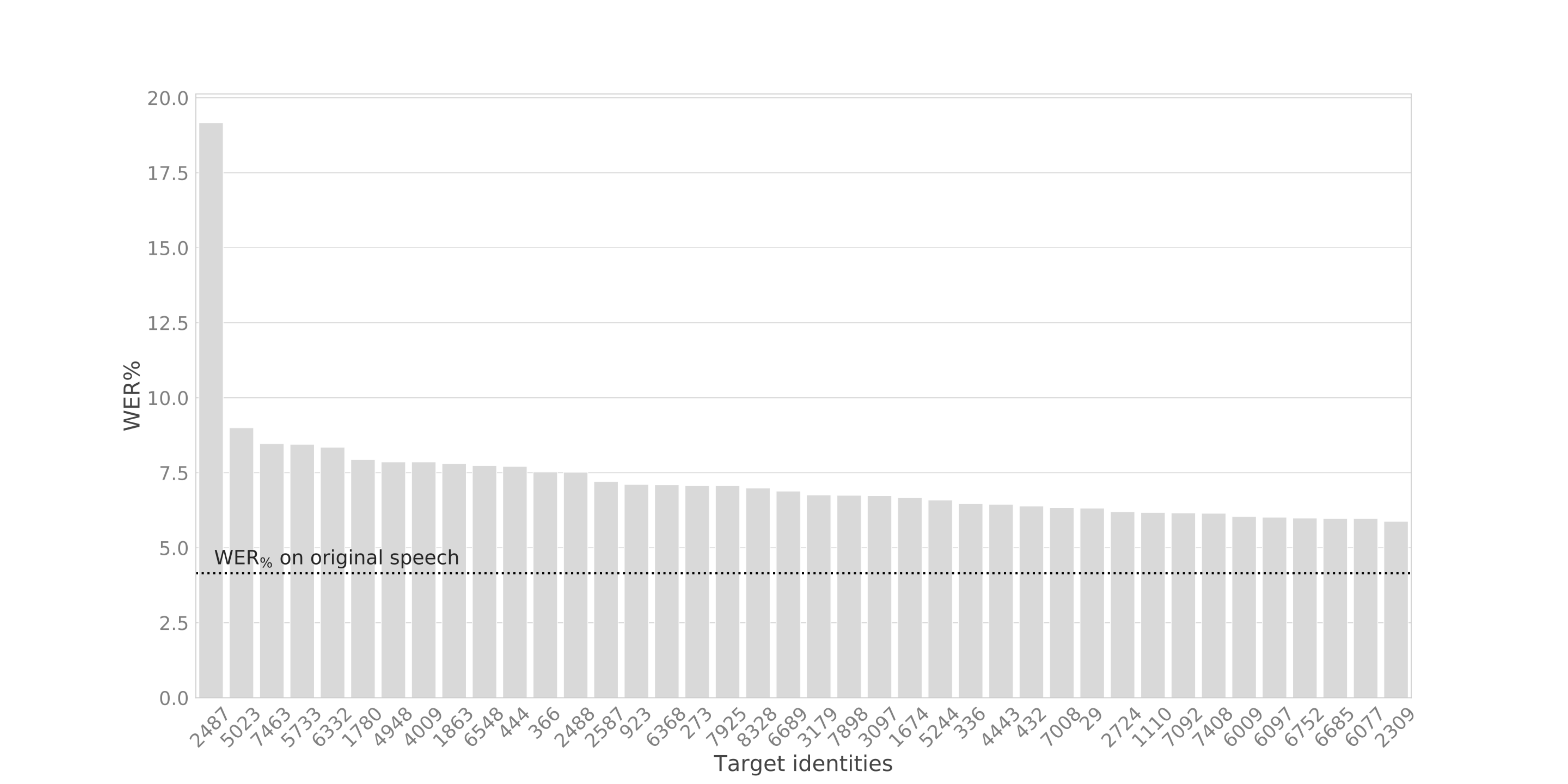}}
  \hspace{2cm}
\end{minipage}
\caption{WER$_\%$ scores obtained by the VoicePrivacy ASR evaluation systems for each of the 40 speakers and on original speech (dotted black line).}

\label{fig:ASR40-res}
\end{figure*}

Across all experiments, we evaluated the utility for each 40 target identities.
We performed the intelligibility test with the pre-trained ASR system of the VoiecPrivacy challenge.
Figure \ref{fig:ASR40-res} shows the utility scores after anonymization considering each of the 40 target identities. On original clean speech signal, the ASR systems score 4.15 WER$_\%$. 
When using the same model (trained on clean data) the overall WER$_\%$ on the anonymized data reaches 7.30 WER$_\%$ (average value over the 40 experiments).
Retraining the ASR system on anonymized speech improves the WER$_\%$ significantly \cite{tomashenkoVoicePrivacy2020Challenge}.
The very high utility loss yielded when using speaker 2487 is due to a generalization issue of the anonymization system (with happens with and without the F0 modification), in this case, non-intelligible speech was generated at the beginning of some segments. 
We conducted an additional test using 100 randomly selected target speaker identities, and were able to find 3 target speaker x-vectors that have the same behavior, with the worse case having a score of 59.37 WER$_\%$. 
Informal listening test reported that the original speech that produced the faulty x-vector contained singing segments. Further analysis needs to be conducted.

\vspace{-2em}

\vspace{-2em}

\section{Conclusion}

\label{sec:conclusion}

In this work, we evaluated and analyzed the x-vector-based speaker anonymization system proposed in \cite{fangSpeakerAnonymizationUsing2019} and F0 extension proposed in \cite{f0_mod_moi} under a white-box attack scenario approach and a \textit{constant} target selection strategy. 
To assess the long-term performance of the voice conversion system and the impact of each of the hyper-parameters (i.e. the x-vector target identity, the F0 linear transformation, and the  random gender selection), the use of a \textit{constant} target selection strategy is beneficial. 
This target selection strategy allows the attacker to have complete knowledge about the system,
under those circumstances models and hyper-parameters design choices can be compared as the best attacker will be used to evaluate the anonymization performance.

The experiments done in \cite{f0_mod_moi,brij_vpc_design} showed that black and grey-box attackers can easily be fooled when enrollment speech and trial speech are anonymized in a different manner, i.e. by selecting the x-vector in different regions, applying the F0 linear transformation and random gender selection.
On the contrary, we observed that neither the x-vector target, F0 transformation, or cross-gender conversion actually help to remove the speaker information from the speech signal.
We showed that regardless of the hyper-parameters used, privacy protection stays the same. 
The detailed analysis showed that a given set of hyper-parameters does not help the anonymization system to better anonymize a given source speaker. Furthermore, we concluded that the anonymization system performance depends on the speaker to anonymize, half of our test speakers did not have their privacy improved.

We raise caution on the privacy evaluation procedure, as we've shown that system performance varies depending on the attacker's knowledge and ASV system used. 
In future work, we plan to evaluate the source of the speaker information leakage that occurs through the phonetic features (PPGs).

\vspace{-1.2em}

\subsubsection{Acknowledgments.}

This work was supported in part by the French National Research Agency under project DEEP-PRIVACY (ANR-18-CE23-0018) and Région Grand Est. Experiments were carried out using the Grid’5000 testbed, supported by a scientific interest group hosted by Inria and including CNRS, RENATER and several Universities as well as other organizations.

%
%
%
\bibliographystyle{splncs04}
\bibliography{mybibliography.bib}
\end{document}